\documentclass[fleqn,usenatbib,useAMS]{mnras}

\usepackage{graphicx}
\usepackage{amsmath}
\usepackage{amssymb}
\usepackage{times}

 \def\be{\begin{equation}}
 \def\ee{\end{equation}}
\def\solmas{{{\rm M}$_\odot$}}
\def\simless{\mathbin{\lower 3pt\hbox
   {$\rlap{\raise 5pt\hbox{$\char'074$}}\mathchar"7218$}}}   
\def\simgreat{\mathbin{\lower 3pt\hbox
   {$\rlap{\raise 5pt\hbox{$\char'076$}}\mathchar"7218$}}}   

\def\solm{{\rm M}_\odot}

\def\ms {M_*}

\def\racc {R_{\rm acc}}

\def\macc {\dot M_*}
\def\menc {M_{\rm enc}}
\def\Rclus {R_*}
\def\Rbh {R_{\rm BH}}
\def\Rtidal {R_{\rm tidal}}
\def\vrel {v_{\rm rel}}

\title[Accretion and the Baryonic Fraction in Galaxies]{Competitive Accretion and the Baryonic Fraction in Galaxies}
\author[I. A. Bonnell]{
 Ian A.~Bonnell\thanks{E-mail: iab1@st-andrews.ac.uk} \\  SUPA, School of Physics \& Astronomy, University of
    St Andrews, North Haugh, St Andrews, Fife KY16 9SS, United Kingdom }
  
\begin{document}

\maketitle

\begin{abstract}  
The baryonic fraction of galaxies is observed to vary with the mass of its dark matter (DM) halo. Low-mass galaxies have low baryonic fractions which increase to a maximum for masses near $10^{12}\ \solm$, and decreases thereafter with increasing galaxy mass. This trend is generally attributed to the action of feedback from star formation at the low end and of active galactic nuclei at the high-mass end.
An alternative is that the baryonic fraction is at least partially due to the ability of galaxies to competitively accrete gas in a group or clustered environment. Most galaxies in a group including those of lower masses, orbit the cluster centre at significant speeds and hence their accretion is limited by a Bondi-Hoyle type process, $\dot{M}_{acc} \propto M_{DM}^2$. In contrast, the few high-mass galaxies reside in the core of the cluster and accrete in a tidal  accretion process, $\dot{M}_{acc} \propto M_{DM}^{2/3}$. These two mechanisms result in a baryonic mass fraction that increases as $M_{DM}$ at low masses and decreases as $M_{DM}^{-1/3}$ at high masses. This model predicts that lower-mass halos in small-N groups should have higher baryonic fractions relative to those in large clusters.

\end{abstract}

\section{Introduction} \label{s:intro}

Galaxy formation is understood in terms of the hierarchical merger of dark matter halos and subsequent infall of baryons to form the luminous components of galaxies \citep{WF1991,LC1993,NFW1995,Cetal2000,Springel2005,DeLB2007,Vogelsberger2014,Bose2023}. The mass of baryons in galaxies is observed to depend on the mass of the dark matter halo, with lower-mass halos having  a smaller fraction of their total mass in stars and gas \citep{Papastergis2012,Zaritsky2023,Dev2025}. The mass fraction in baryons increases with dark matter halo mass up to masses of $M_{\rm DM} \approx 10^{12}$ \solmas\ and decreases at higher halo masses. 

This variation in the efficiency of dark matter halos to accumulate baryons is often attributed to the effects of feedback that remove baryons from the dark matter halos. Star formation and feedback are primary internal processes that can regulate galaxy evolution \citep{DT08,SchEA15,ReyRaposo2017}, suggesting that feedback from supernova could be responsible for removing baryons in lower mass, and hence lower escape velocity halos \citep{Efstathiou2000,Kay2002,MW03,SEA06,DVS2008,DVS2012,KEA14,Nelson2019,Tollet2019,Mina2021}. Feedback from active galactic nucleii (AGN) is significant in higher mass galaxies suggesting that AGN feedback can remove baryons from these galaxies \citep{Silk2010,Mead2010,Martizzi2012,Wright2020,Cui2021}. The challenge with feedback is that it tends to find weakpoints in the surrounding environment through which to escape while leaving the bulk of the mass unaffected \citep{Dale2005,Dale2012,RogersPittard2013,Kortgen2016,Lucas2020,Lau2025}.

While feedback is an important process and can play an important role in galactic evolution including removing baryons, it is worthwhile exploring what other physical processes can affect the efficiency of baryon accretion in dark matter halos.  Many to most galaxies are found to be in groups or clusters \citep{Makarov2011,Courtois2013,Kourkchi2017,Lambert2020} with local estimates having only 11\% of galaxies in isolation  and a further 10\% in double or triple systems \citep{ArgudoF2015}. 
In small groups to large clusters,  gravitational forces from neighbours and the full cluster can affect the dynamics. 
Accretion of baryons can be affected and limited by these competing forces as the gravitational radius of influence depends on the mass of individual objects. In such a scenario, dark matter halos would compete to accrete from the baryonic reservoir present in the cluster. 

Competitive accretion in star formation has been advanced as a potential mechanism the explain the stellar initial mass function \citep{Zin1982,BBCP2001,BLZ2007}. Accretion in a cluster environment is also a prime mechanism to form high-mass stars and result in a mass-segregated system where the highest-mass stars form in the deepest part of the cluster potential \citep{BVB2004}. The requirements for competitive accretion to occur is that the stars form in groups and clusters, and that there exists a shared reservoir of gas to accrete \citep{BB2006}. The process for galaxy baryonic accretion is similar in that galaxies are generally in groups or clusters and acquire their baryonic components by infall into dark matter halos. The main difference is that the accretion does not significantly increase the overall mass of the objects as that is determined by the dark matter halo merger process. 

This paper investigates the role that competitive accretion in galaxy clusters can play in determining the efficiency of baryon acquisition by dark matter halos. \S~2 reviews the physics of competitive accretion while \S~3 details how this is adapted to the case of baryon accretion in galaxy clusters. The results are presented in \S~4 and conclusions in \S~5.

\section{Competitive Accretion}

Competitive accretion in stellar clusters has been advanced as a method to explain the origin of the stellar initial mass function \citep{Zin1982,BBCP1997,BBCP2001, BCBP2001, BBV2003,BLZ2007}. The basic concept is that stars form in a clustered environment with a communal gravitational potential and a shared reservoir of gas from which they compete to accrete. The competition arises through the gravitational influence of each object and how this depends on its mass. The two possibilities are that 1) the accretion is limited by the  kinetic motions of the gas and star, and 2) that the accretion is limited by the tidal forces due to all the other objects in the cluster.  

An object can accrete gas at a rate given by
\begin{equation}
\macc \approx \pi \rho \vrel \racc^2,
\end{equation}
where $\rho$ is the gas density,  and $\vrel$ is the relative velocity of the gas. The  accretion radius,  $R_{\rm acc}$, is the radius at which gas is bound to the accreting object considering both the relative kinetic energy of the gas and the tidal forces from the full cluster.  Considering only the relative kinetic motions results in the classical Bondi-Hoyle radius 
\begin{equation}
\Rbh \approx 2G\ms/(\vrel^2 + c_s^2).
\end{equation}
For roughly spherical systems, the tidal radius can be approximated as being due to the enclosed mass ($\menc$) inside the object's position in the cluster $\Rclus$,  
 a tidal radius is given by 
\begin{equation}
\Rtidal \approx 0.5  \left({\ms / \menc}\right)^{1\over 3} \Rclus. 
\end{equation}

Numerical simulations of accretion in stellar clusters \citep{BBCP2001} has shown that the majority of stars residing outside the core of the cluster have relatively high velocities resulting in the Bondi-Hoyle radius being smaller and hence determinant for accretion rates. In contrast stars in the core of the cluster have relatively low velocities and hence the tidal radius is the limiting radius and hence determines the accretion rates. These objects have the highest accretion rates and end up forming the most massive stars \citep{BVB2004,BB2006,BSCB2011}. Lower-mass objects formed (at the local Jeans mass) in the outer parts of the cluster accrete little from the intracluster gas and hence lose out in this competition and remain with low masses reproducing the peak and the breadth of the stellar IMF \citep{BCB2006,BB2006,BSCB2011,KBS2025}.

\section{Competitive accretion for dark matter halos}

Dark matter halos residing in clusters and groups can accrete baryons from their shared communal reservoir. In what follows, I develop a simple scenario where all baryons are assumed to accrete onto already formed dark matter halos in such a system. The primary difference with the stellar IMF scenario above is that the overall galaxy mass is given by the mass of the dark matter halo. The accretion of baryons does not increase the mass of the dark matter halo, and hence the accretion rate remains nearly constant as baryons accumulate in the dark matter halo. The two cases for accretion in groups is that in the core of a group or cluster the lower velocities present imply high Bondi-Hoyle radii such that tidal processes are the dominant physics that limits accretion.  The mass of baryons acquired over a time $t_{\rm acc}$ is then

\begin{equation}
M_{\rm baryon} \approx \macc t_{\rm acc},
\end{equation}
with the lower-mass dark matter halos having accretion rates given by 
\begin{equation}
\macc \approx \pi \rho {\left( G M_{DM}\right)^2 \over \left (\vrel^2 + c_s^2\right)^{3/2}}.
\end{equation}
The final baryonic fraction then scales with halo mass as
\begin{equation}
f_{\rm DM} = {M_{\rm baryon}\over M_{\rm DM}} \propto M_{\rm DM}.
\end{equation}
The higher-mass halos are generally located in the cluster cores and thus have accretion rates 
\begin{equation}
\macc \approx \pi \rho \vrel \left( {M_{DM} \over {\menc}} \right)^{2/3} \Rclus^2,
\end{equation}
and hence a baryonic mass fraction that scales as
\begin{equation}
f_{\rm DM} = {M_{\rm baryon}\over M_{\rm DM}} \propto {M_{\rm DM}}^{-1/3}.
\end{equation}
These scalings are promising as they roughly agree with the observed distributions, yet we need a model that includes the halo mass function and dependencies within the galaxy clusters.

\section{Baryon accretion in galaxy clusters}

In order to assess the potential role for accretion in determining the baryonic fraction of galaxies, We use a simple model for how baryonic accretion occurs in galaxy clusters. It neglects any baryons included in the initial halo formation or any subsequent baryon removal through stripping or ejection.
The model contains 1000 clusters  based on observed properties  \citep{Makarov2011,Kourkchi2017,Sing2025}, with each  cluster composed of between 5 and 2000 dark matter halos, following a log-uniform $dN_{\rm gals} \propto N_{\rm gals}^{-1}$ distribution. The individual halos are randomly chosen from a dark matter halo mass function \citep{Driver2022}, with minimum and maximum masses of $10^{9.5}$ and $10^{14.5}$ \solmas, respectively:
\begin{equation}
\Phi \left(\log_{10} \left({M / \solm}\right)\right) \propto \left( {M \over M_*}\right)^{\alpha + 1} \exp{ - \left({M \over M_*}\right)^{\beta}},
\end{equation}
where $M_*=10^{14.43} \solm$, $\alpha =-1.85$, and $\beta=0.77$.

 The clusters have a fiducial radius of $R_{\rm clust} = 1$ Mpc, with near-uniform density cores  inside 0.05$R_{\rm clust}$. The cores contain five per cent of the cluster members, with a minimum of 2 and a maximum of 25.  The dark matter halos in the core are chosen to be the most-massive members in the cluster as is commonly found in clusters. The rest of the halos are located at distances outside the core in either a mass-segregated or a non-mass segregated distribution.  The total baryonic mass is taken to be 10 per cent of the dark matter mass in the cluster. The cluster cores are relatively small as they represent where the mass density is near uniform.

\begin{figure}
	\begin{center}
	\includegraphics[scale=0.55]{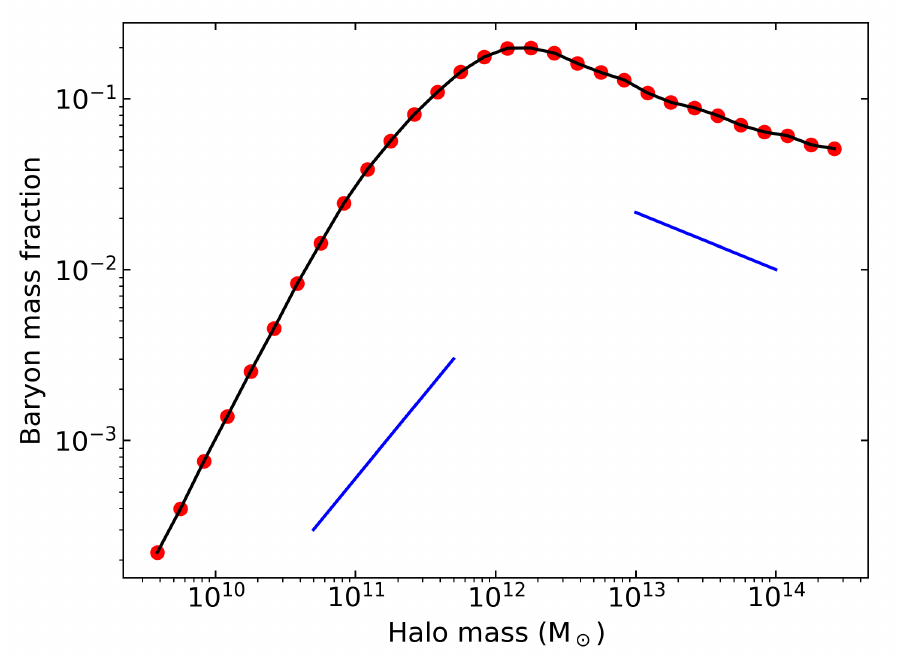}
	\caption{The average baryonic mass fraction is plotted against dark matter halo mass. The accretion rates are calculated for mass-segregated cluster with density profiles $\rho\propto r^{-1}$. The two lines show the expected relationships,  for Bondi-Hoyle accretion (low-mass halos), $f_{\rm baryon}  \propto  M_{\rm DM}$,  and tidal-lobe accretion (higher-mass halos), $f_{\rm baryon}  \propto  {(M_{\rm DM}})^{-1/3}$. The data are for cluster gas density profiles $\rho\propto r^{-1}$ and cluster number functions that are $dN_{\rm gals} \propto N_{\rm gals}^{-1}$.}
	\label{figgalacc1}
	\end{center}
	\vspace{-0.5cm}
\end{figure}

For dark matter halos in the core, their accretion radii are calculated as the minimum of their tidal and Bondi-Hoyle radii (as above) and then they accrete accordingly with uniform velocities given by the core velocity dispersion and uniform core densities. Similarly, halos outwith the core have their tidal and Bondi-Hoyle radii evaluated, with a ubiquitous smaller Bondi-Hoyle radii, which is then used to evaluate the accretion along with a uniform cluster velocity dispersion ($v_{\rm clust} = \sqrt{2G M_{\rm clust} / R_{\rm clust}}$) and a power-law density $\rho = \rho_{\rm core} \left(R/R_{\rm core}\right)^{-\gamma}$ with $1 \le \gamma \le 2$ \citep{Vikhlinin2006,Meneghetti2014}. Halos accrete over two cluster crossing times ($t_{\rm acc} = 2 t_{\rm cross}$) to attain a final baryonic mass and hence a baryonic mass fraction as plotted in Figure~\ref{figgalacc1} for cluster gas density profiles $\rho\propto r^{-1}$ and cluster number functions that are $dN_{\rm gals} \propto N_{\rm gals}^{-1}$. 

The galaxy halo mass function is such that halos near or above $10^{12}$ \solmas\ are always amongst the most massive objects in the cluster and hence reside in the core. Lower-mass halos are generally found outwith the cluster core unless the cluster or group consists of a small number of objects. Figure~\ref{figgalacc1} shows the mean baryonic fraction of galaxies as a function of dark matter halo mass. As halos with $M_{\rm DM} \ge 10^{12}$ \solmas\ are located in the core of the cluster, they accrete via tidal-lobe accretion and hence result in baryonic mass fractions which decreases as $f_{\rm baryon}  \propto  {(M_{\rm DM}})^{-1/3}$.

The majority of lower-mass halos ($M_{\rm DM} \le 10^{12}$ \solmas) are located outside the core and hence accrete via Bondi-Hoyle accretion. This results in an average baryonic mass fraction that increases with mass at a somewhat steeper rate than  $f_{\rm baryon}  \propto  M_{\rm DM}$, the  pure Bondi-Hoyle expectation discussed above. The complications are due to 1) a cluster gas density profile that decreases with radius and a degree of mass-segregation along with 2) a number of cases where lower mass halos are still the most massive member of a small-N group. The former can result in baryonic mass fractions as steep as $f_{\rm baryon}  \propto  {(M_{\rm DM})}^{\nu}$ with $\nu = 1 + \gamma$ and $\gamma$ is the gas density profile power-law index (see above). The latter results in halos with high $f_{\rm DM}$  as they are in the core of low-N groups. Together this results in a wide spread of $f_{\rm DM}$ at low halo masses and an average $f_{\rm DM} \propto (M_{\rm DM})^{1+\epsilon}$ with $\epsilon \approx 0.5$.

\begin{figure*}
	\begin{center}
	\hbox{\includegraphics[scale=0.4]{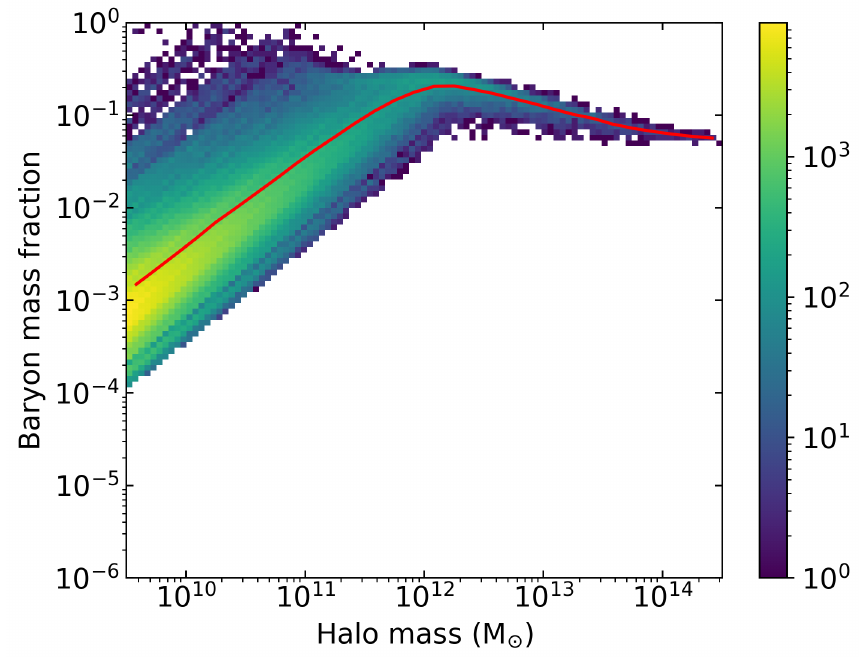},\includegraphics[scale=0.4]{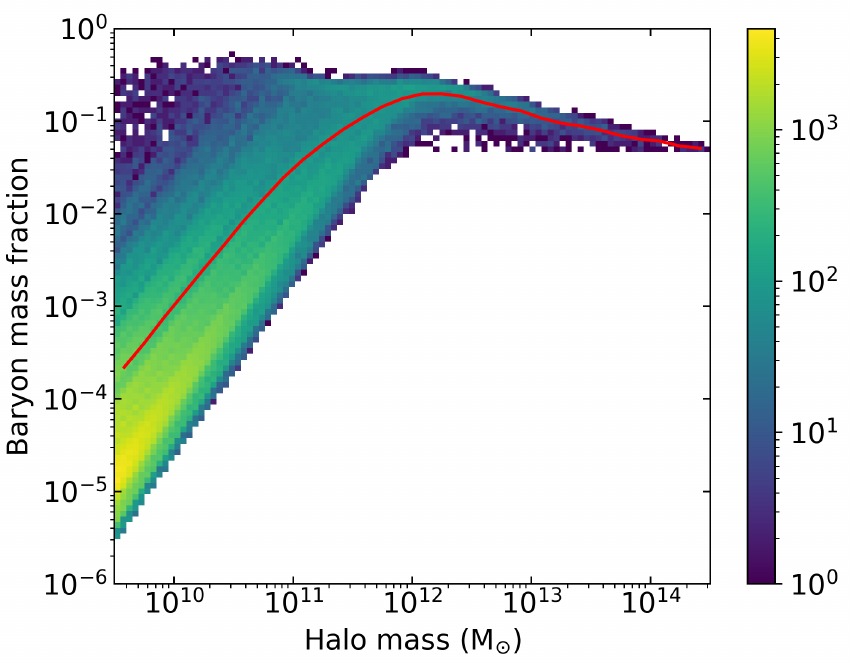},\includegraphics[scale=0.4]{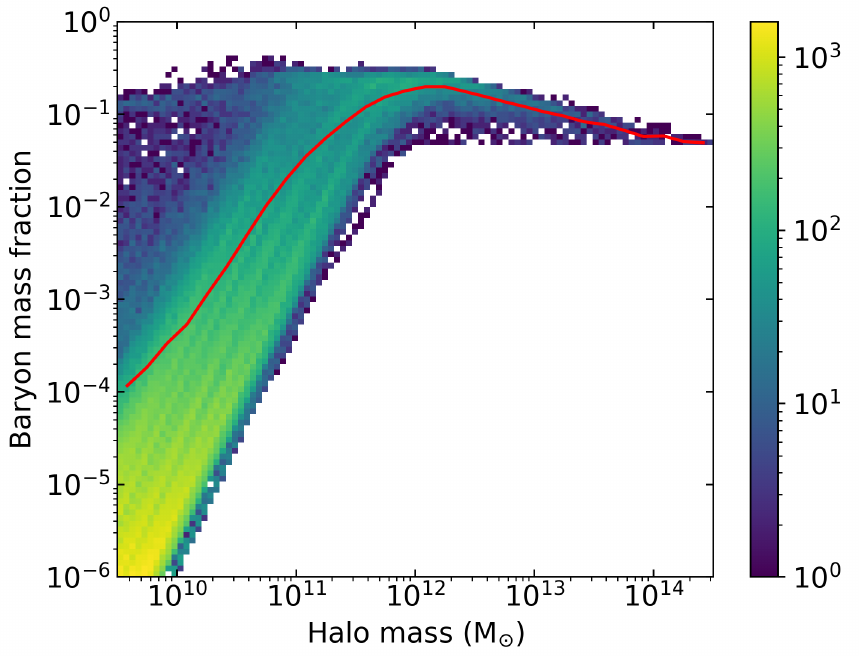}}
	\caption{The baryonic fractions are plotted as a function of dark matter halo mass as a 2-D histogram and the average baryonic fractions as red lines. The panels represent different cluster parameters with the left-hand panel having uniform gas densities in the cluster and a flat cluster $dN_{\rm gals} \propto N_{\rm gals}^0$ distribution. The middle panel has a $\rho\propto r^{-1}$ and an $dN_{\rm gals} \propto N_{\rm gals}^{-1}$ cluster distribution while the right hand panel has a $\rho\propto r^{-2}$ and $dN_{\rm gals} \propto N_{\rm gals}^{-2}$ cluster distribution.}
	\label{figgalsacc2d}
	\end{center}
	\vspace{-0.95cm}
\end{figure*}

\subsection{Model Parameters}

Figure~\ref{figgalacc1} plots the mean baryonic fraction for dark matter halos for clusters that are mass-segregated, have gas densities
$\rho \propto r^{-1}$ and follow galaxy number distributions of $dN_{\rm gals} \propto N_{\rm gals}^{-1}$. The baryonic fraction varies significantly at a given halo mass depending on the cluster properties and especially at lower masses. The middle panel of Figure~\ref{figgalsacc2d} plots a 2-D histogram of the full baryonic mass fractions across the full 1000-cluster simulation. At low halo masses we see a large variation in resultant baryonic fractions depending on the cluster sizes with low-$N_{\rm gals} $ clusters likely to have a lower-mass halo as its most massive member. In these cases the central dark matter halo accretes more than its neighbours and has a relatively high baryonic mass fraction. In high-$N_{\rm gals}$ clusters, lower-mass halos have accretion rates that are lower by $f_{\rm baryon}  \propto  \rho M_{\rm DM}$ and as $\rho \propto r^{-1}$ in the mass-segregated system, this  results in a lower-limit of $f_{\rm baryon}  \propto  (M_{\rm DM})^2$. Taken together this produces a mean baryonic mass function at low $M_{\rm DM}$ that has $f_{\rm baryon}  \propto  (M_{\rm DM})^{1.2}$.  The high-mass $M_{\rm DM}$ follows a $f_{\rm baryon}  \propto  (M_{\rm DM})^{-1/3}$ distribution in all three panels of Figure~\ref{figgalsacc2d}.

Figure~\ref{figgalsacc2d} explores how the resultant baryonic fraction depends on our chosen parameters, the cluster density profile and the distribution of the number of clusters with given galaxy populations \citep{Vikhlinin2006,Meneghetti2014}. The left-hand panel shows the baryonic mass fractions for dark matter halos in clusters where the gas density is taken to be uniform and the distribution of clusters with given $N_{\rm gals}$ galaxies is flat $dN_{\rm gals} \propto N_{\rm gals}^0$. In this case we have a mean $f_{\rm baryon}  \propto  M_{\rm DM}$ and a variation around this where the mean and mode are similar. Most halos are in high-$N_{\rm gals}$ clusters with lower-mass halos accreting as $\macc \propto (M_{\rm DM})^2$. 

The right-hand panel of Figure~\ref{figgalsacc2d} explores the other extreme where the cluster number density distribution is $dN_{\rm gals} \propto N_{\rm gals}^{-2}$ and the cluster gas follows a steep profile of $\rho\propto r^{-2}$. In this case, the lower-mass halos in high $N_{\rm gals}$ clusters follow a steep baryonic mass fraction with $f_{\rm baryon}  \propto  (M_{\rm DM})^3$. This is offset by the larger number of low $N_{\rm gals}$ clusters in which lower-mass halos are more likely to be amongst the more massive members and hence accrete following tidal lobe accretion. The relatively high number density at low $M_{\rm DM}$ and high $f_{\rm baryon}$ is apparent in this panel.  Other model parameters have been explored  including the proportion of halos in the core, the core radius and the degree of mass segregation outside the halo. These all have small affects that are within the results for the models presented in Figure~\ref{figgalsacc2d}.

\section{Discussion}
The results of the simple toy model presented in this paper show that accretion onto dark matter halos can conceivably produce baryonic mass fractions that are similar to what are observed.   
 The underlying assumption is that most galaxies are in groups or clusters during this accretion phase. The fraction of galaxies in clusters and groups is uncertain due to completeness and the definition of group or cluster size. Published estimates show of order 50 to 80\% are in  larger groups or clusters or in the process of infalling into the cluster potential \citep{Courtois2013,Makarov2011,Lambert2020}. \cite{ArgudoF2015} find only 10\% of the galaxies in the nearby Universe are truly isolated with a further 10\% in double or triple systems. Competitive accretion can operate in small-N systems \citep{Bonnell1997} such that the process described here may be applicable to a large majority of galaxies. Galaxies that are not found in groups and clusters would neither gain baryons from the higher accretion rates in the cluster cores, nor lose the competition for baryons in the periphery of the group or cluster. Such galaxies, presumably of lower halo masses, would likely have a higher baryonic fraction than their counterparts that are in groups and clusters. 

In addition to producing steeply increasing baryon fraction at low halo masses and a shallow decreasing baryonic fraction at high halo masses,  the model results in a peak near halo masses of $10^{12}$ \solmas\ due to the halo mass function \citep{Driver2022} and having higher-mass halos predominantly located in cluster cores.  The model's greatest attraction is its simplicity in relying entirely on gravitational processes. This simplicity harbours unknown challenges in terms of the missing physics  such as feedback from AGN and supernova. A complete picture also requires exploration of the dynamics of cluster formation and the interior dynamics where accretion occurs. This is well-beyond the scope of this paper and its intention to highlight the role that competitive accretion can play in determining the baryonic mass fraction of galaxies.

 The resulting baryonic fractions from this model are simply due to baryonic accretion onto already formed dark matter halos. In reality, baryons can accumulate at all stages of the hierarchical merger process from the formation of the first low-mass halos through the subsequent mergers  and dynamics in groups and clusters. This adds a significant degree of complexity to this simple picture, with processes such as tidal and ram-pressure stripping \citep{Merrit1983,Abadi1999,Mayer2006,Read2006,McCarthy2008,Baselli2022} and feedback \citep{Silk2010,DVS2012} all likely to play a role in removing baryons, although tidal stripping may play a larger role in removing mass from the dark matter halos rather than the more concentrated baryons \citep{Smith2016}.  Feedback during the accretion phase is most likely to have an important role in limiting accretion rather than removing baryons from deep inside the halos. 

The hierarchical merger of dark matter halos implies the most massive halos are just accumulations of many lower mass halos. In the absence of baryonic accretion during this process, the final baryonic mass fraction would simply reflect the primordial baryonic mass fraction in each small-mass halo. Numerical simulations of galaxy and large scale structure formation \citep{Vogelsberger2014,Artale2016,Matthee2016,Bose2019,Cui2021} highlight the importance of baryonic accretion onto the dark matter halos  and can be used to explore how this accretion depends on local properties such as locations in the groups, relative velocities and tidal effects \citep{Pillepich2018,Springel2018}. The complexity present in full dynamical simulations will undoubtably be more involved than the simple picture presented here but accretion rates can be analysed in respect to the predicted physical processes of tidal and Bondi-Hoyle accretion. 

What we have highlighted in this paper is that accretion in clusters is likely to follow two regimes that produce marked differences in the amount of baryonic gas that can be accreted onto dark matter halos. The model finds a robust $f_{\rm baryon}  \propto  (M_{\rm DM})^{-1/3}$ in all models explored independent of the cluster properties. This is due to the tidal-lobe accretion which dominates in the core of clusters. Lower-mass halos have mean baryonic mass fractions that decrease below $M_{\rm DM}\approx 10^{12}$\solmas\ but with significant variations. The exact slope of $f_{\rm baryon}  \propto  (M_{\rm DM})^{\mu}$ varies but with a mean in the ballpark of $1\le \mu\le 3$ with most likely values near $1 \le \mu \le 1.5$.   A prediction for the model is that lower-mass halos in small-N groups, or in isolation,  should have higher baryonic mass fractions. Conversely, lower-mass halos in larger groups, or where they are significantly less massive than most group halo, should have very low baryonic mass fractions.

\section{Conclusions}
This paper presents a simple model for baryonic accretion onto dark matter halos in galaxy groups and clusters in order to explain the observed baryonic fraction as a function of halo mass. The model uses simple prescriptions for accretion in groups whereby more massive halos reside in the cores of clusters and accrete via a tidal process. In contrast, lower-mass halos reside outside the cores and accrete via a Bondi-Hoyle process determined by the virial velocity in the cluster. These two processes result in an increasing baryonic mass fraction at low halo masses with $f_{\rm baryon}  \propto  (M_{\rm DM})^{\mu}$ with $\mu \approx 1 {\rm\  to\  } 1.5$, a peak baryonic fraction for $M_{\rm DM} \approx 10^{12}$\solmas\ and a decreasing $f_{\rm baryon}  \propto  (M_{\rm DM})^{1/3}$ for high-mass halos. Low-mass halos have significant dispersion in their $f_{\rm baryon}$ as they can be members of large-N clusters or equally small-N groups of galaxies in which they are more likely to be the most massive members.   A prediction of this model is that lower-mass halos in small-N groups or in isolation should have larger baryon fractions than those in large groups and clusters.

The model presented here is a simple toy-model and a complete model would require the incorporation of the halo formation and merger process  in the context of a cosmological model. Feedback from supernova and AGN would also be important additional physics to include in order to definitively ascertain the role of competitive accretion in determining the baryonic fraction in galaxies.

\section*{Acknowledgements}
IAB would like to that Keith Horne, Jim Pringle, Paul Clark and Kat Klos and the referee for comments and discussions which helped improve the paper.

\section*{Data Availability}
The data generated in this work can be made available upon request to the author.

\end{document}